\documentclass[%
reprint,
 amsmath,amssymb,
 aps,
]{revtex4-2}

\usepackage{graphicx}
\usepackage{dcolumn}
\usepackage{bm}

\usepackage{slashed}
\usepackage{mathtools}
\usepackage{graphicx}
\usepackage{braket}
\usepackage{caption}
\usepackage{subcaption}
\newcommand{\beq}{\begin{equation}}
\newcommand{\eeq}{\end{equation}}
\newcommand{\bea}{\begin{eqnarray}}
\newcommand{\eea}{\end{eqnarray}}

\def\ba{\begin{array}}
\def\ea{\end{array}}
\def\sQ3{\widetilde{Q}_3}
\def\sU3{\widetilde{U}_3}
\def\sD3{\widetilde{D}_3}
\newcommand{\fbinv}{\text{fb}$^{-1}$}

\begin{document}


\title{Probing Invisible Decay of $Z^\prime$ at Muon Collider with Topological Data Analysis and Machine Learning}

\author{Jyotiranjan Beuria}%
 \email{jyotiranjan.beuria@gmail.com}
\affiliation{IKS Research Center, ISS Delhi, Delhi, India}

\date{\today}

\begin{abstract}
We explore the use of topological data analysis (TDA) combined with machine learning for 
discriminating standard model backgrounds from the invisible decay of the $Z^\prime$ boson associated with monophoton emission at a 3 TeV muon collider. 
Reconstructed events are mapped into a six-dimensional kinematic space and aggregated into 
bags of events, from which persistent homology is used to extract Betti number distributions. 
Within the Multiple Instance Learning paradigm, classifiers trained on these topological 
descriptors demonstrate significantly improved classification accuracy compared to the conventional 
ML approaches based on event-wise kinematic inputs. We also draw exclusion contours at 95\% CL in the $(m_{Z^\prime}, m_\chi)$ parameter space, 
highlighting the potential of topological features to extend the discovery reach of future 
collider experiments.
\end{abstract}

\maketitle


\section{Introduction}

Extensions of the Standard Model (SM) often predict the existence of an additional neutral 
gauge boson, generically denoted as $Z^\prime$ \cite{langacker2009physics}. 
Such states arise in a wide range of frameworks, including grand unified theories \cite{robinett1982prospects,hewett1989low,langacker2009physics}, string-inspired constructions \cite{cvetivc1996new}, and phenomenological models with an extra $U(1)$ gauge symmetry \cite{arun2022search,su2024complex, barik2025discovering,das2025testing}. 
Depending on the structure of the couplings, a $Z^\prime$ can decay both to visible SM 
particles and to hidden-sector states, the latter giving rise to invisible final states 
associated with dark matter candidates~\cite{alves2015dark,inan2022search,wang2025prospects,feng2025dark}. 
Probing invisible $Z^\prime$ decays is therefore an important task in collider phenomenology, 
as it provides a direct connection between gauge extensions of the SM and the dark sector.


The Large Hadron Collider (LHC) has placed stringent bounds on visible $Z^\prime$ 
channels, particularly through dilepton and dijet resonance searches~\cite{ATLAS:2019erb,CMS:2018ipm,sirunyan2020multi}. These analyses exclude 
new gauge bosons up to the multi-TeV scale, depending on the model assumptions for couplings 
to quarks and leptons. However, probing invisible decays of 
a $Z^\prime$ at the LHC is significantly more challenging. The dominant search strategy relies 
on monojet or monophoton plus missing transverse energy ($\slashed{E}_T$) signatures, which 
suffer from large QCD backgrounds and systematic uncertainties in the reconstruction of missing 
energy \cite{alves2014dark,alves2015dark}. As a result, LHC sensitivity to invisible $Z^\prime$ decays is considerably weaker than 
for visible final states, leaving an important gap in our understanding of possible dark sector 
connections.  

Future high-energy lepton colliders offer a much cleaner experimental environment to 
study such processes. The absence of QCD backgrounds and the precisely known initial state 
allow for more sensitive probes of invisible channels. Proposed facilities such as the 
ILC (International Linear Collider)~\cite{baer2013international}, 
CLIC (Compact Linear Collider)~\cite{abramowicz2017higgs}, and 
FCC-ee/CEPC (circular electron-positron colliders)~\cite{abada2019fcc,agapov2022future} 
are designed primarily as Higgs and electroweak factories, excelling in precision measurements 
but operating at center-of-mass energies below a few TeV. By contrast, a 
muon collider proposed by the International Muon Collider Collaboration (IMCC) uniquely combines the electroweak cleanliness of lepton collisions with 
center-of-mass energies in the multi-TeV range~\cite{delahaye2019muon,long2021muon,accettura2023towards}. At $\sqrt{s} = 3$~TeV, a muon collider can directly produce a heavy $Z^\prime$ well 
beyond the LHC reach, and thanks to the direct coupling of muons to the new gauge boson, 
production rates can remain sizable even for suppressed couplings. This makes the muon collider 
an especially powerful machine to explore invisible decays and hidden-sector dynamics of a 
$Z^\prime$.

A particularly powerful feature of muon colliders in this context is the 
radiative return mechanism~\cite{Binner:1999bt}. 
When the collider operates at a fixed center-of-mass energy above the $Z^\prime$ pole, 
the emission of an energetic photon from the initial state effectively lowers the invariant 
mass of the muon pair, allowing the $Z^\prime$ resonance to be accessed. 
This process, leading to $\mu^+ \mu^- \to \gamma Z^\prime$, is especially important for invisible 
decays of the $Z^\prime$, as the final state consists of a hard photon recoiling against 
missing energy. The resulting mono-photon signature is clean, with the primary backgrounds 
coming from SM processes such as $\mu^+ \mu^- \to \gamma \nu \bar{\nu}$. In this work, we consider the production of $\gamma Z^\prime$ and subsequent invisible decay of $Z^\prime$ to a pair of invisible Dirac fermions.

In such a scenario, an effective separation of signal and background requires exploiting both the kinematic signatures and the recoil structure of the final state. In $\gamma\slashed{E}_T$ searches, the dominant SM background 
$\mu^+\mu^- \to \gamma \nu \bar{\nu}$ mimics the invisible $Z^\prime$ signal and is almost
irreducible given the larger production cross section of the SM process. Simple cut-and-count strategies capture only coarse differences, 
discarding correlated kinematic information essential for discrimination.
Thus, modern machine learning (ML) techniques \cite{radovic2018machine,schwartz2021modern,karagiorgi2022machine} are crucial for mono-photon searches, 
since they can exploit subtle correlations in photon kinematics, recoil mass, and event 
shapes that cut-based analyses might miss. By combining multiple observables, ML classifiers 
significantly improve the sensitivity to an invisible $Z^\prime$ over the smooth the SM 
$\mu^+ \mu^- \to \gamma \nu \bar{\nu}$ background.

The well motivated traditional approach in collider phenomenology relies on event-level kinematic distributions such as photon energy, 
transverse momentum, and missing transverse energy, and other such kinematic variables. While effective, these observables may not fully exploit the collective patterns present in 
collider observations. In this work, we propose a complementary approach that leverages topological data analysis (TDA)~\cite{chazal2021introduction,beuria2023persistent,beuria2024intrinsic}, 
which extracts higher-order geometric and topological information from ensembles of events. 
By representing reconstructed events as point clouds in a multidimensional kinematic space 
and computing their persistent homology, we obtain the so-called Betti curves that capture global 
structural differences between signal and background events. 

These features are then used within machine learning classifiers with
multiple instance learning (MIL) framework~\cite{carbonneau2018multiple} to enhance signal-background separation. In the MIL framework, data are grouped into bags of instances, with labels assigned at the bag level rather than per instance. In collider physics, this allows one to exploit collective patterns in sets of reconstructed events, enhancing sensitivity when single-event observables are not sufficiently discriminating \cite{beuria2024intrinsic}. We call this approach, TDA-ML in contrast to the traditional ML approach. As we demonstrate, this TDA-ML approach provides significant improvements in classification accuracy. We also explore the resulting exclusion limits in search of $Z^\prime$ at 95\% CL.

The paper is organized as follows. In section~\ref{sec:framework}, we introduce the $Z^\prime$ model and outline the radiative return leading to hard photon recoil. We also introduce some basics of topological data analysis.
We discuss the collider simulation in section~\ref{sec:collider} and the machine learning pipeline in section~\ref{sec:ml}. In section~\ref{sec:results}, we present the outcomes of our detailed simulations, including classification accuracies and projected exclusion limits, with particular emphasis on the impact of radiative return. Finally, we summarize our findings and discuss future directions in section~\ref{sec:conclusion}.

\section{The Theoretical Framework}
\label{sec:framework}
\subsection{The Model}

The interaction structure of the model originates from gauge kinetic mixing 
between the Standard Model $U(1)_Y$ and an additional $U(1)'$ gauge symmetry. 
After diagonalization through electroweak symmetry breaking (EWSB), the new spin-1 mediator, $Z^\prime$ acquires effective couplings to the SM leptons and the dark matter fermion $\chi$. 
The relevant interaction Lagrangian in this simplified model is:
\begin{align}
\mathcal{L} \;\supset\;
&- e \, \bar{\mu} \gamma^\mu \mu \, A_\mu \notag \\
&+ \bar{\mu}\,\gamma^\mu 
   \left( g_{\mu} + a_{\mu}\gamma^5 \right) \mu \, Z'_\mu \notag \\
&+ \bar{\chi} \, \gamma^\mu 
   \left( g_{\chi} + a_{\chi}\gamma^5 \right) \chi \, Z'_\mu .
\end{align}
The Lagrangian above consists of three terms. 
The first corresponds to the standard QED coupling of the muon to the photon, 
which enables initial state radiation. 
The second describes the interaction of the muon with the new spin-1 mediator 
$Z'_\mu$ through vector and axial-vector couplings $g_{\mu}$ and $a_{\mu}$. 
The third term introduces the interaction between the Dirac fermion dark matter 
$\chi$ and the mediator $Z'_\mu$, with couplings $g_{\chi}$ and $a_{\chi}$. 
Together, these interactions govern the process 
$\mu^+ \mu^- \to Z' \gamma$ followed by the decay $Z' \to \bar{\chi}\chi$, 
leading to a final state with a hard photon accompanied by missing energy 
carried by the dark matter pair.

At leading order (LO), neglecting $m_\mu$, the cross section for 
$\mu^+\mu^- \to Z' \gamma$ scales as
\begin{equation}
\sigma_{\rm LO} \sim 
\frac{\alpha (g_\mu^2+a_\mu^2)}{s}
\left(1-\frac{M_{Z'}^2}{s}\right)
\Big[\ln\!\left(\tfrac{s}{m_\mu^2}\right)-1\Big],
\end{equation}
which vanishes when $M_{Z'}^2\simeq s$. Including ISR, the photon spectrum factorizes into a radiator function and the 
reduced cross section at $\hat{s}=s(1-x_\gamma)$, with 
$x_\gamma=2E_\gamma/\sqrt{s}$,
\begin{align}
\frac{d\sigma}{dx_\gamma} \simeq 
\frac{\alpha}{\pi}\,
\frac{1+(1-x_\gamma)^2}{x_\gamma}\,
\ln\!\left(\tfrac{s}{m_\mu^2}\right)\,
\sigma_{\mu^+\mu^-\to Z'}(\hat{s}),
\end{align}
where
\begin{align}
\sigma_{\mu^+\mu^-\to Z'}(\hat{s}) \propto
\frac{(g_\mu^2+a_\mu^2)\,\hat{s}}
{(\hat{s}-M_{Z'}^2)^2+M_{Z'}^2\Gamma_{Z'}^2}.
\end{align}

The Breit-Wigner factor leads to \emph{radiative return}: a hard photon lowers 
the effective invariant mass so that $\hat{s}\simeq M_{Z'}^2$, producing a 
resonant photon at
\begin{equation}
E_\gamma^* \simeq \frac{s-M_{Z'}^2}{2\sqrt{s}}.
\label{eq:estar}
\end{equation}

\subsection{Topological Data Analysis}
Topological data analysis (TDA) provides a set of tools to extract 
topological invariants from discrete data and study their stability under 
continuous deformations. The central idea is to construct a family of 
simplicial complexes from a point cloud and track the evolution of their 
homological features as a function of a scale parameter, also known as the filtration scale ($\epsilon$). 

Given a finite metric space $\{x_i\}$, one defines the Vietoris-Rips (VR) 
complex $\mathrm{VR}_\epsilon$ at scale $\epsilon$ as the abstract simplicial 
complex containing a $k$-simplex whenever all pairwise distances among its 
$k+1$ vertices are smaller than $\epsilon$. As $\epsilon$ increases, 
$\mathrm{VR}_\epsilon$ forms a nested sequence of complexes
\begin{equation}
\mathrm{VR}_{\epsilon_1} \subseteq \mathrm{VR}_{\epsilon_2} 
\subseteq \cdots \subseteq \mathrm{VR}_{\epsilon_m},
\end{equation}
which induces a sequence of homology groups. Persistent homology 
captures the birth and death of homology classes across this filtration, 
summarized by persistence diagrams or barcodes. 

The $k$-th homology group, denoted as $H_k$, is the quotient group, representing cycles modulo boundaries \cite{carlsson2009topology}. Mathematically, it is expressed as:
\begin{equation}
    H_k = \frac{Z_k}{B_k} = \frac{\text{ker}(\partial_k)}{\text{im}(\partial_{k+1})}. 
\label{eq:hk}
\end{equation}
Here, $H_k(K)$ is the quotient vector space whose generators are given by $k$-cycles that are not boundaries of any $(k+1)$-simplices. The rank of the $k$-th homology group is the $k$-th Betti number $\beta_k$, 
counting independent $k$-dimensional holes: $\beta_0$ corresponds to 
connected components, $\beta_1$ to cycles or loops, and $\beta_2$ to voids. 
By plotting $\beta_k$ as a function of $\epsilon$, one obtains Betti curves as the dimension of the $k$-th homology group, $H_k(\mathrm{VR}_\epsilon)$. In other words, we have
\begin{equation}
\beta_k : \epsilon \mapsto \dim H_k(\mathrm{VR}_\epsilon),
\end{equation}
which provides compact one-dimensional summaries of topological structure. 

Persistence summaries such as Betti curves or persistence landscapes are robust descriptors with stability guarantees under perturbations. Thus, they have been successfully employed as features in machine learning pipelines, offering a way to incorporate global and multiscale structural information that is not accessible to purely statistical methods. For more details, the reader is advised to refer to some standard reviews on topological data analysis \cite{carlsson2009topology, murugan2019introduction, carlsson2020topological,chazal2021introduction}.

\section{Collider Simulation}
\label{sec:collider}
We study resonant $Z^\prime$ production in association with mono-photon at a $\sqrt{s}=3~\text{TeV}$ muon collider at 100 \fbinv, 
followed by invisible decays into dark matter particles. The characteristic collider 
signature is a hard photon from initial state radiation (ISR) recoiling against large 
missing transverse energy ($\slashed{E}_T$). The dominant Standard Model (SM) background 
arises from neutrino pair production with an associated photon, 
$\mu^+\mu^- \to \nu \bar{\nu} \gamma$. In our simulations, we include both 
$\nu_e\bar{\nu}_e\gamma$ and $\nu_\mu\bar{\nu}_\mu\gamma$ channels. 

Event generation is performed at leading order (LO) with 
\texttt{MadGraph5\_aMC@NLO v3.6.3}~\cite{Alwall:2014hca,Frederix:2018nkq}. For both the SM and the BSM processes, we have kept a minimum $p_T$ for the ISR photon to be $150$ GeV during the parton-level event generation. For the $\gamma+\slashed{E}_T$ signal, which involves no colored final states, the parton-level events could be passed directly to fast detector simulation. We also note that while the hard ISR photon is responsible for the radiative return, which is generated at the matrix-element level, \texttt{Pythia v8.315}~\cite{bierlich2022comprehensive} can additionally 
simulate softer ISR/FSR photons, which may slightly modify the photon 
distributions, but do not affect the primary radiative return peak. Thus, we choose to pass the parton-level events through \texttt{Pythia v8.315}. Both SM and the BSM model cross sections are normalized with an NLO $K$-factor of 1.2. We also consider a flat factor of 10\% in systematic uncertainties for the background processes. A fast detector simulation is performed using \texttt{Delphes~3.5.0}~\cite{deFavereau:2013fsa}, 
with photon reconstruction implemented within the acceptance of a muon collider detector. 
We employ the default muon collider configuration provided in \texttt{Delphes~3.5.0}.

We define a set of baseline cuts common to all signal regions, and then distinguish between two photon recoil windows around $E_\gamma^*$.
We apply the following baseline selection to isolate high-energy photon recoil events:
\begin{equation}
\begin{aligned}
p_T^\gamma > 200~\text{GeV}, \\
\slashed{E}_T > 200~\text{GeV}, \\
|\eta_\gamma| < \eta_{\rm max} \;(\theta > 10^\circ) \approx 2.44, \\
\Delta\phi(\gamma,\slashed{E}_T) > 2.5 .
\end{aligned}
\end{equation}

Two signal regions are then defined by the photon energy window around $E_\gamma^*$ (see equation~\ref{eq:estar}):
\begin{itemize}
    \item SR-A: $|E_\gamma - E_\gamma^*| \geq 0$, 
    \item SR-B: $|E_\gamma - E_\gamma^*| < 100~\text{GeV}$.
\end{itemize}

SR-A is the generic region without specifically focusing on the photon recoil energy $E_\gamma^*$. For SR-B, we select events with photon energies within a $\pm 100~\text{GeV}$ window 
around the expected peak value $E_\gamma^*$. $|\eta_\gamma|$ is restricted by requiring the photon polar angle 
$\theta > 10^\circ$, corresponding to the detector acceptance outside the 
shielding cone, and $\Delta\phi(\gamma,\slashed{E}_T)$ denotes the azimuthal separation between the photon and the missing transverse momentum. These cuts suppress the large background from soft or collinear photon emission in $\nu \bar{\nu} \gamma$ production, while maintaining sensitivity to the hard photon recoil spectrum expected from resonant 
$Z^\prime$ production.

We consider $g_\mu=0.02$ and $g_\mu=0.1$ for the $Z^\prime$ model. The other two independent model parameters, namely,  $m_{Z^\prime}$ and $m_{\chi}$, are varied from 400 GeV to 3000 GeV and 0 to 1500 GeV in steps of 100 GeV such that $2 m_{\chi} \leq m_{Z^\prime}$. For each of these parameter points, we simulate $10^5$ parton-level events and process them through showering and fast detector simulation. The same protocol is also followed for $2\times 10^6$ parton-level events for the SM background. We further extract six kinematic features from the reconstructed events at the detector for further processing of our TDA feature extraction. These six features are as follows:
\begin{itemize}
  \item $E_\gamma$ : Photon energy (GeV),
  \item $p_T^\gamma$ : Photon transverse momentum (GeV),
  \item $\eta_\gamma$ : Photon pseudorapidity,
  \item $\phi_\gamma$ : Photon azimuthal angle,
  \item $\slashed{E}_T$ : Missing transverse energy (GeV),
  \item $\Delta\phi(\slashed{E}_T, \gamma)$ : Azimuthal angle difference between $\slashed{E}_T$ and the photon.
\end{itemize}

To explore the utility of machine learning backed by topological data analysis, we consider two strategies for our analysis. In one case, we use TDA features to train an ML model, and in another case, we train an ML model directly on the six kinematic features. Henceforward, we denote the first method as ``TDA-ML" and the second one, ``ML". The details of the machine learning framework is described below.
\section{Machine Learning Framework}
\label{sec:ml}
The implemented framework compares three classifiers under a uniform experimental design: 
Support Vector Machine (SVM), Random Forest (RF), and a one-dimensional Convolutional Neural Network (1D-CNN). 
Two datasets are provided, corresponding to  signal ($X_{\text{bsm}}$) and background ($X_{\text{sm}}$). 
These are vertically stacked to form the full feature matrix
\[
X = \begin{bmatrix} X_{\text{bsm}} \\ X_{\text{sm}} \end{bmatrix}, 
\quad 
y = [1,\ldots,1,0,\ldots,0],
\]
where class label $1$ denotes signal and $0$ denotes background. The evaluation protocol is $5$-fold stratified cross-validation, ensuring class balance in each fold. The model with the highest mean accuracy is selected, and the corresponding confusion matrix is utilized to obtain signal and background efficiency.

The binary confusion matrix can be written as shown in Table \ref{tab:confmatrix}.
\begin{table}[!ht]
\centering
\setlength{\tabcolsep}{6pt} 
\renewcommand{\arraystretch}{1.0} 
\small 
\[
\begin{array}{c|c|c}
 & \text{Predicted Signal} & \text{Predicted Background} \\
\hline
\text{True Signal} & TP & FN \\
\hline
\text{True Background} & FP & TN \\
\end{array}
\]
\caption{Binary confusion matrix. 
$TP$: true positives, $FN$: false negatives, 
$FP$: false positives, $TN$: true negatives.}
\label{tab:confmatrix}
\end{table}
Thus, the signal selection efficiency is defined as the fraction of true signal events that survive the selection, i.e., 
\[
\epsilon_{\text{sig}}^\text{ml} = \frac{TP}{TP+FN},
\]
which corresponds to the true positive rate (recall) in machine learning. On the background side, one can define the background efficiency as
\[
\epsilon_{\text{bkg}}^\text{ml} = \frac{TN}{FP+TN},
\]
the fraction of true background events identified as background. Thus, for a given integrated luminosity $\mathcal{L}$ and SM background (BSM) process cross section $\sigma_{\text{bkg}}$ ($\sigma_{\text{sig}}$) with kinematic-cut efficiency $\epsilon_{\text{bkg}}^\text{cut}$ ($\epsilon_{\text{sig}}^\text{cut}$), the number of background and signal events left for statistics is as follows:
\begin{align}
    s &= \sigma_{\text{sig}} \times \mathcal{L} \times \epsilon_{\text{sig}}^\text{cut} \times \epsilon_{\text{sig}}^\text{ml},\\
    b &= \sigma_{\text{bkg}} \times \mathcal{L} \times \epsilon_{\text{bkg}}^\text{cut} \times \epsilon_{\text{bkg}}^\text{ml}.
\end{align}
Once we have the number of signal, $s$, and background, $b$ events, we can define the significance level as follows:
\begin{equation}
Z = \sqrt{ \, 2 \left[ (s+b)\,\ln\!\left(1+\frac{s}{b}\right) - s \right] }.
\label{eq:asimov}
\end{equation}
The 95\% CL corresponds to $Z\approx1.64$. Later, we will make use of this formalism while drawing exclusion contours at 95\% CL.

Now we delve into another important issue related to the dataset preparation. The dimension of the feature vector is 300 for the TDA-ML framework and 6 for the ML framework. In the case of TDA-ML, we are considering Betti number distribution with 100 bins across three homology dimensions ($H_k$, $k=0,1,2$). On the contrary, the ML framework utilises six kinematic features mentioned earlier.

For the SVM and RF classifiers, the pipeline explicitly includes two preprocessing steps: 
(i) feature scaling using \texttt{StandardScaler}, and 
(ii) dimensionality reduction using Principal Component Analysis (PCA) with $6$ retained components in the case of the TDA-ML framework. 
This reduces redundancy and stabilizes optimization. 
The SVM employs an RBF kernel with $C=1.0$, while the RF uses $200$ estimators with parallel execution enabled.

\begin{figure*}[!ht]
  \centering
  \includegraphics[width=\textwidth]{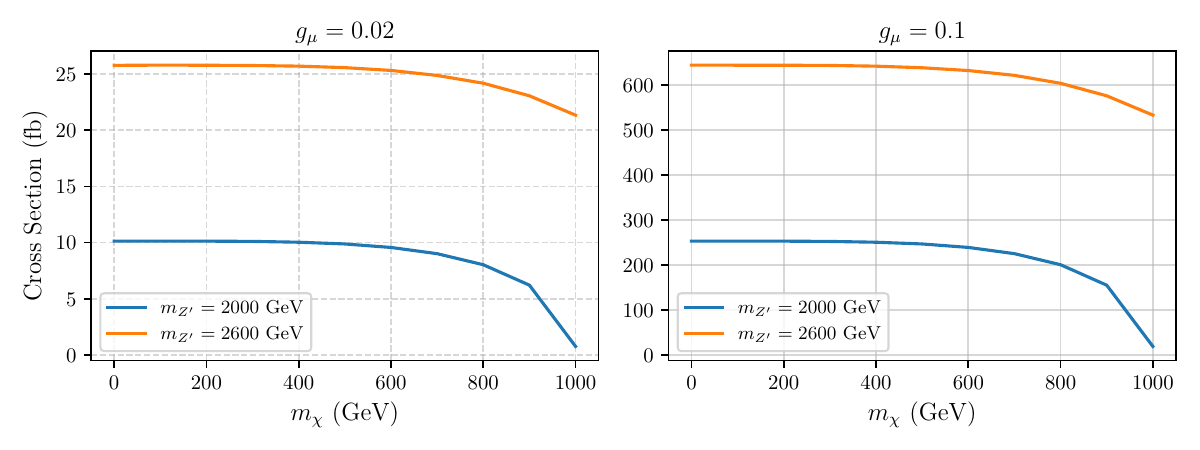}
  \caption{Variation of production cross section of $\mu^+ \mu^- \rightarrow \gamma \bar{\chi}\chi$ with the mass of fermion dark matter $\chi$ for coupling strengths $g_\mu=0.02$ and $g_\mu=0.1$ and   $m_{Z^\prime}=2000$ GeV and $2600$ GeV. The heightened cross section for $m_{Z^\prime}=2600$ GeV corresponds to the mechanism of radiative return.}
  \label{fig:bsm_xsection}
\end{figure*}

The CNN is implemented manually, with a repeated cross-validation loop. Each fold involves standardization of the training data, reshaping of features into $(n_{\text{features}}, 1)$ format, and construction of a Conv1D architecture: two convolutional layers (32 and 64 filters, kernel size 3), global max pooling, a fully connected dense layer of 64 units with ReLU, dropout with a rate $0.3$, and a final sigmoid output. Early stopping is used to prevent overfitting, monitoring validation loss with patience $5$. Predictions are thresholded at $0.5$ to yield binary class outputs. 

In the case of the TDA-ML framework, we need to form the point cloud so as to construct a simplicial complex. We denote the feature vector having six kinematic features as the position vector in a six-dimensional space. In order to form a point cloud from reconstructed events at the detector, we prepare bags of 100 events for both the SM background and the BSM signal. Thus, in the case of the TDA-ML framework, training and testing are actually done on the bags of signal and background. The signal selection and background rejection efficiencies are considered for further processing.

In the case of the TDA-ML framework, the natural training unit is not a single reconstructed event but a bag of events, since the topological features are extracted from point clouds formed by aggregating multiple events in the six-dimensional kinematic space. Each bag of events inherits a class label (SM or BSM), while the individual events within the bag are not necessarily discriminative on their own. Thus, the TDA-ML falls under the category of weekly supervised learning. This setting is well captured and justified by the paradigm of Multiple Instance Learning (MIL), where labels are attached to bags of instances rather than to individual instances~\cite{carbonneau2018multiple}. In this sense, our use of signal selection and background rejection efficiencies from the bag level classification is consistent with the MIL formalism and justifies the adoption of MIL-inspired approaches for training and evaluating classifiers on TDA-derived features.

\section{Results and Discussion}
\label{sec:results}

Figure~\ref{fig:bsm_xsection} shows the variation of the production cross section $\mu^+\mu^- \rightarrow \gamma \bar{\chi}\chi$ as a function of the dark matter mass $m_\chi$ for two large values of the mediator mass, namely, $m_{Z^\prime}=2000$~GeV and $2600$~GeV, and for couplings $g_\mu = 0.02$ and $g_\mu = 0.1$. As expected, the cross section decreases with increasing $m_\chi$, reflecting the reduction in available phase space for heavy dark matter production. The comparison of the two couplings highlights the strong dependence of the rate on $g_\mu$, 
with the $g_\mu=0.1$ case yielding cross sections nearly an order of magnitude larger than for $g_\mu=0.02$. 
Furthermore, the enhancement observed for the heavier mediator mass $m_{Z^\prime}=2600$~GeV 
is a manifestation of the radiative return mechanism, which allows the collider energy 
to effectively match the mediator resonance through photon emission, thereby amplifying 
the production rate. 
At large $m_\chi$, the suppression is more severe for $m_{Z^\prime}=2000$~GeV compared to 
$m_{Z^\prime}=2600$~GeV, leading to visibly steeper cross-section falloff in the lower mediator 
scenario.

\begin{figure*}[!ht]
  \centering
  \includegraphics[width=\textwidth]{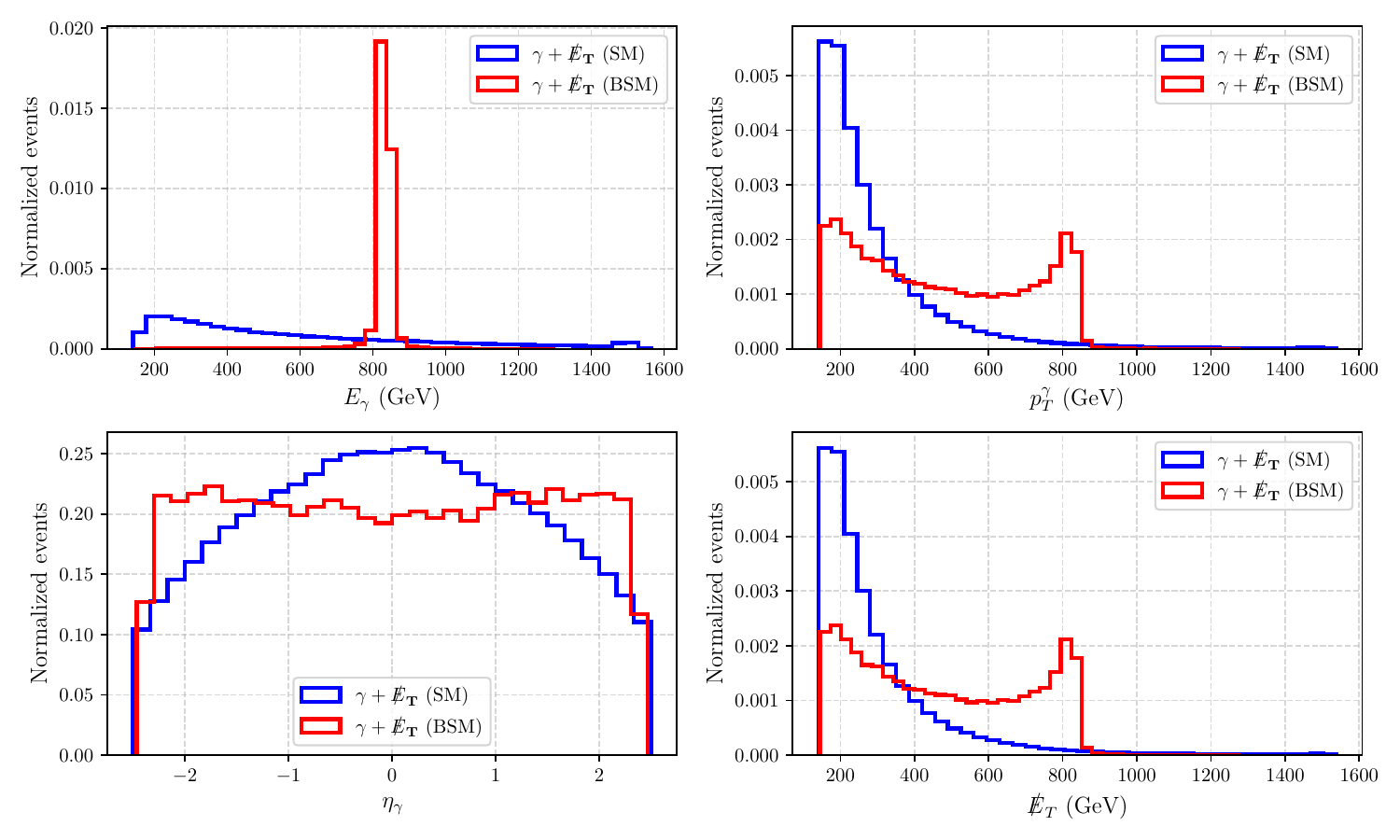}
  \caption{Kinematic distributions of the signal and the background process for $m_{Z^\prime}=2000$ GeV and $m_{\chi}=500$ GeV at $g_\mu=0.02$.}
  \label{fig:kinematic_dist}
\end{figure*}

\begin{figure*}[!ht]
  \centering
 \includegraphics[width=0.75\textwidth]{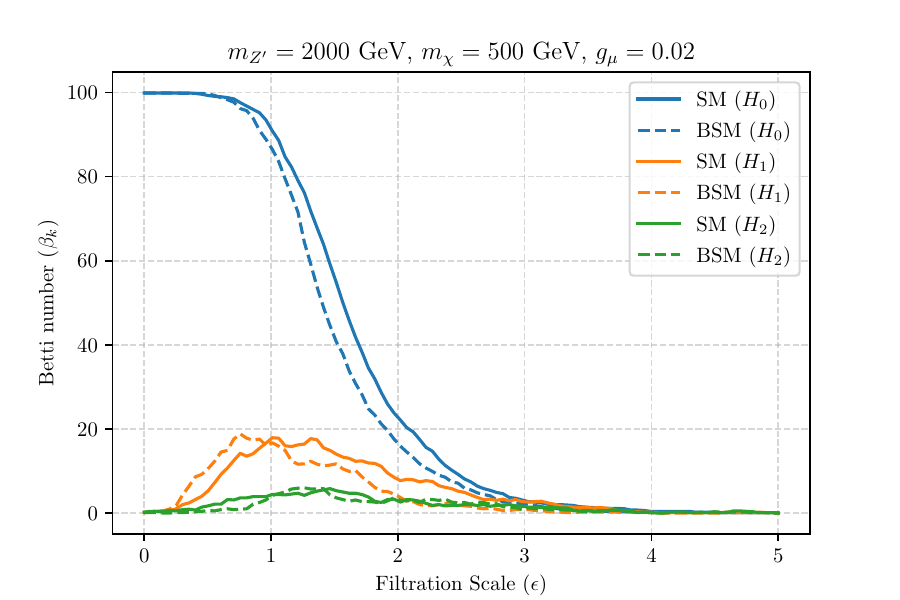}
  \caption{Betti number distributions, also known as Betti curves of the signal and the background process across filtration scale $\epsilon$ for $m_{Z^\prime}=2000$ GeV and $m_{\chi}=500$ GeV at $g_\mu=0.02$.}
  \label{fig:betti_curves}
\end{figure*}

\begin{figure*}[!ht]
  \centering
  \subfloat[]{\includegraphics[width=0.48\textwidth]{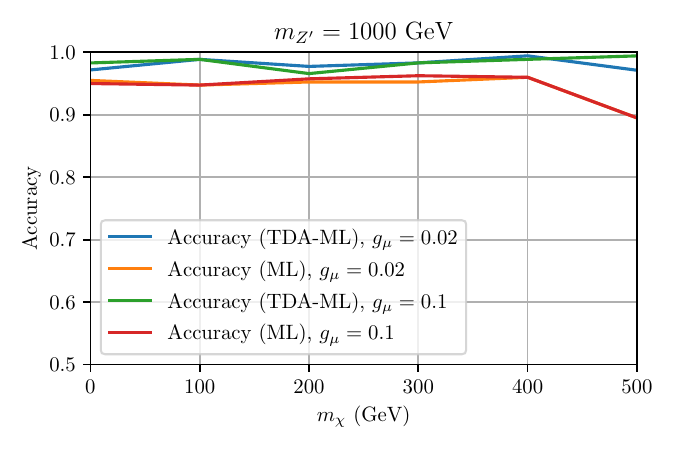}}
  \subfloat[]{\includegraphics[width=0.48\textwidth]{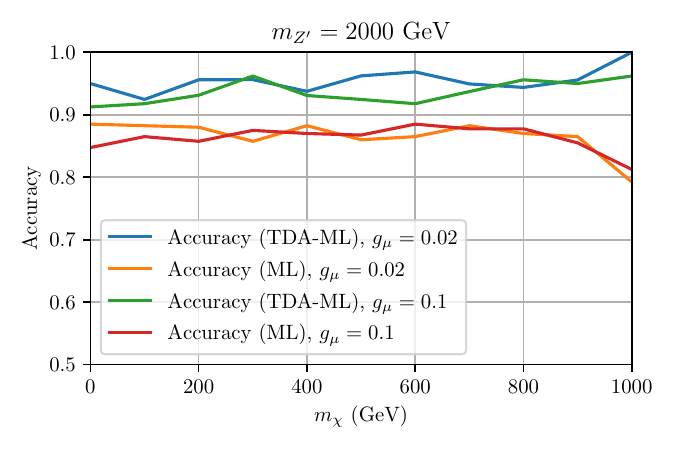}}
  \caption{Classification accuracy as a function of (a) $m_\chi$ for $m_{Z^\prime}=1000$~GeV and (b) $m_{Z^\prime}=2000$~GeV, comparing conventional ML with TDA-ML. The TDA-ML approach based on bag-level topological features achieves consistently higher accuracy across parameter space.}
  \label{fig:acc}
\end{figure*}

Figure~\ref{fig:kinematic_dist} presents the normalized kinematic distributions of the photon energy $E_\gamma$, transverse momentum $p_T^\gamma$, pseudorapidity $\eta_\gamma$, and missing transverse energy $\slashed{E}_T$ for the benchmark point $m_{Z^\prime}=2000$~GeV and $m_{\chi}=500$~GeV. A clear distinction between the SM background and the BSM signal is observed in $E_\gamma$ and $p_T^\gamma$: while the background exhibits broad spectra, the signal shows a sharp peak around $E_\gamma \sim 800$~GeV and $p_T^\gamma \sim 700$~GeV, reflecting the kinematic constraints of dark matter production through the mediator. The $\eta_\gamma$ distribution remains relatively central for both cases, though with a slightly flatter profile for the signal. In contrast, the $\slashed{E}_T$ distribution for the signal is shifted toward higher values compared to the background, consistent with the presence of massive invisible particles in the final state. These differences demonstrate the discriminating power of photon and missing energy observables in separating BSM scenarios from the SM expectation.

\begin{figure*}[!ht]
  \centering
  \subfloat[]{\includegraphics[width=0.5\textwidth]{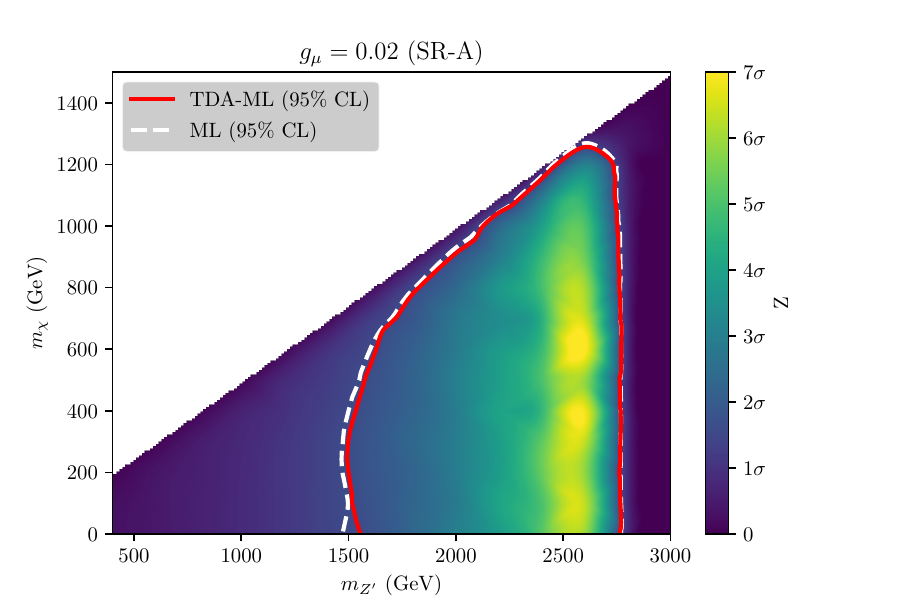}}
  \subfloat[]{\includegraphics[width=0.5\textwidth]{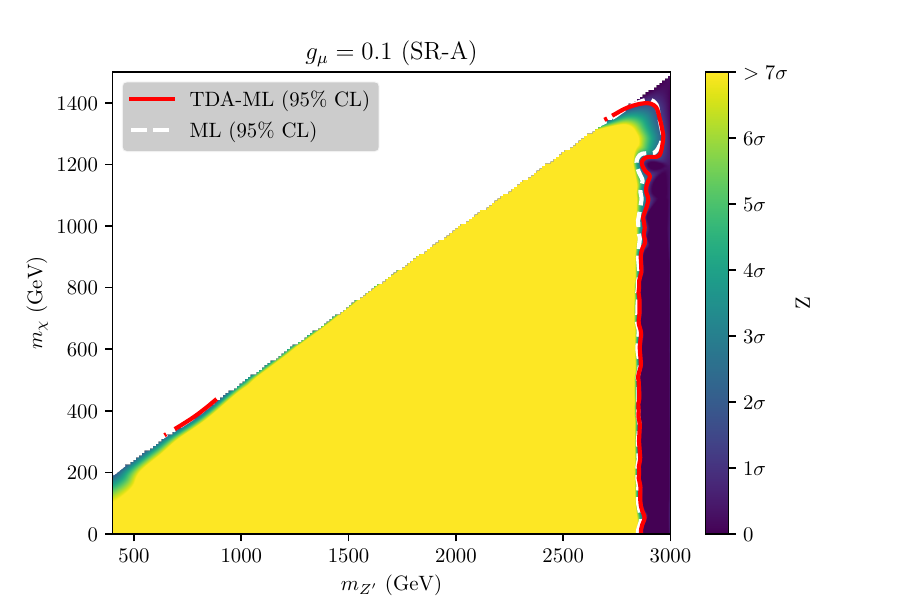}}
  \vskip 0.1in
   \subfloat[]{\includegraphics[width=0.5\textwidth]{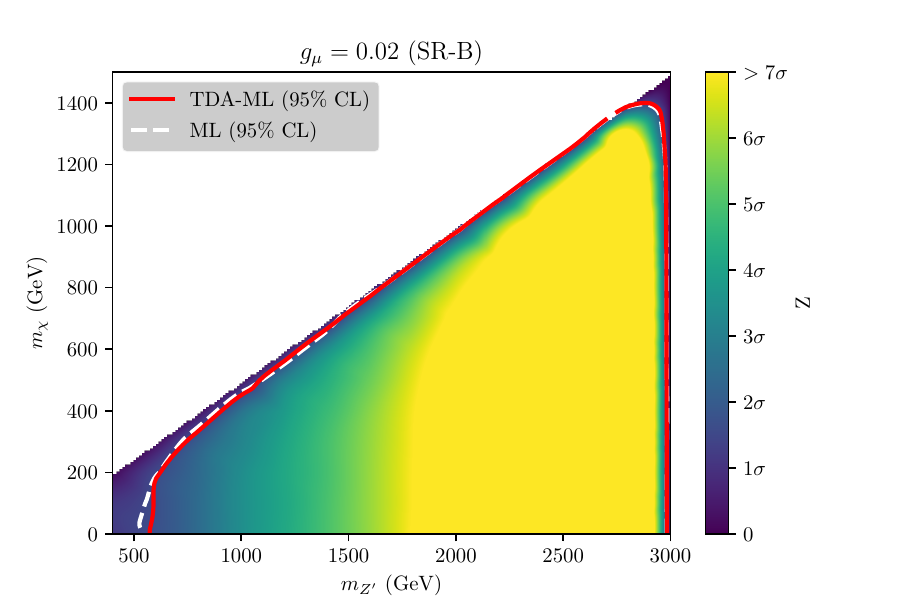}}
  \subfloat[]{\includegraphics[width=0.5\textwidth]{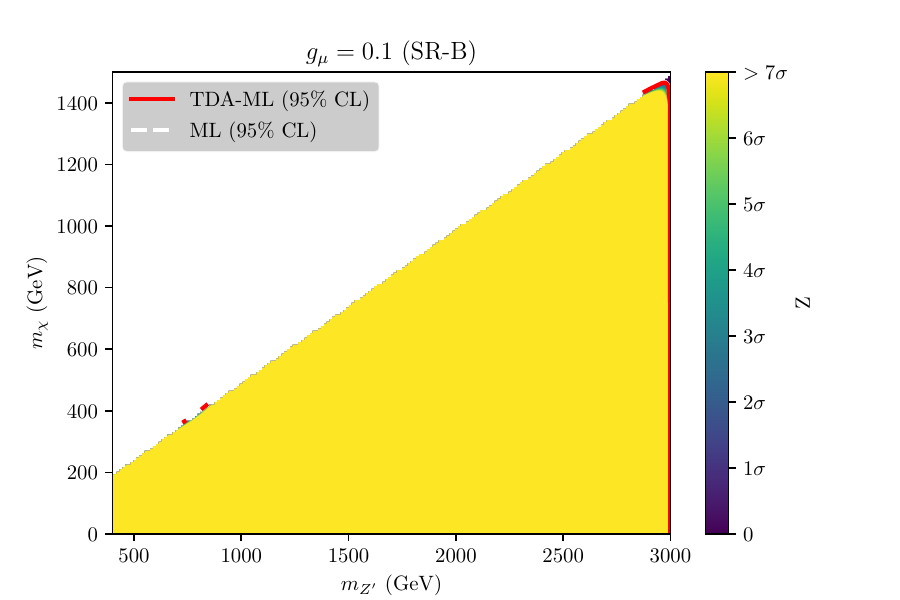}}
\caption{Significance $Z$ in the $(m_{Z^\prime}, m_\chi)$ plane with 95\% CL exclusion contours at 100 \fbinv from TDA-ML (red line) and ML (white dashed line). The top row corresponds to signal region SR-A, while the bottom row corresponds to signal region SR-B, where an additional recoil mass window for the photon is imposed. }
  \label{fig:exclusion_contours}
\end{figure*}

Figure~\ref{fig:betti_curves} shows the Betti number distributions for the benchmark point $m_{Z^\prime}=2000$~GeV, $m_{\chi}=500$~GeV, and $g_\mu=0.02$, comparing the SM background and BSM signal across homology dimensions $H_0$, $H_1$, and $H_2$. The $H_0$ curves, which correspond to connected components, start near the number of points in the cloud and decrease with the filtration scale as components merge; here, the signal exhibits a slightly faster decay than the background, indicating differences in clustering properties. For $H_1$, associated with loops, the BSM case shows a modest enhancement over the SM, particularly around intermediate scales, reflecting the presence of additional cycle structures in the topology. The $H_2$ curves, which capture voids, are overall smaller in magnitude but still reveal mild deviations between the signal and background. These topological differences across homology dimensions provide complementary discriminating features that can be exploited in the TDA-ML framework.

Figure~\ref{fig:acc} shows the signal-background classification accuracy as a function of the dark matter mass $m_\chi$ for two benchmark mediator masses, $m_{Z^\prime}=1000$~GeV and $m_{Z^\prime}=2000$~GeV, and for couplings $g_\mu=0.02$ and $g_\mu=0.1$. In both cases, the TDA-ML framework consistently outperforms the conventional ML approach across the full $m_\chi$ range. For $m_{Z^\prime}=1000$~GeV, the classification accuracy remains close to unity for TDA-ML, with only mild variations as $m_\chi$ increases, while the standard ML method shows a visible drop, particularly at larger couplings where the background overlap is more pronounced. Similarly, for $m_{Z^\prime}=2000$~GeV, TDA-ML maintains a stable accuracy above $0.9$ even at higher $m_\chi$, whereas the usual ML method exhibits a stronger degradation, with accuracies falling toward $0.8$ in the challenging parameter region ($m_{Z^\prime} \approx 2m_\chi$). 

The superior performance of TDA-ML can be attributed to the fact that persistent homology captures the global topological structure of point clouds formed by aggregating reconstructed events. This goes beyond the discrimination power of raw kinematic features, which may exhibit significant overlap between SM and BSM processes on an event-by-event basis. By instead analyzing bags of events, TDA effectively leverages collective differences in clustering, cycles, and voids across the signal and background ensembles. From a machine learning perspective, this is naturally aligned with the Multiple Instance Learning (MIL) paradigm \cite{carbonneau2018multiple}, where classification is performed at the bag level rather than on individual instances. In this way, even if single events are not strongly discriminative, the bag-level topological summary still carries a clear signature of the underlying physics process. The stability of the TDA-ML accuracy curves across parameter space, therefore, demonstrates the robustness of the method and highlights its potential as a powerful complement to traditional ML classifiers in collider phenomenology.

Figure~\ref{fig:exclusion_contours} presents the final exclusion sensitivity in the $(m_{Z^\prime}, m_\chi)$ plane for representative values of the coupling $g_\mu$, shown separately for the low- (left) and high- (right) coupling scenarios. The top row corresponds to signal region SR-A, while the bottom row corresponds to signal region SR-B, where an additional recoil mass window for the photon is imposed. The color map illustrates the expected Asimov significance (see equation \ref{eq:asimov}), and the overlaid contours indicate the $95\%$ CL exclusion limits obtained using the conventional ML framework (white dashed line) and the proposed TDA-ML framework (red solid line). In contrast to more complex search channels where multiple final states are present, the results here demonstrate that TDA-ML and ML perform very similarly in terms of exclusion contours, though they differ at the classification accuracy level. We observe only marginal differences in the excluded parameter space. This similarity can also be attributed to the rather simplistic kinematic phase space of the process under consideration, which limits the extent to which topological descriptors extracted from persistent homology can provide complementary discriminatory power. Nevertheless, small gains from TDA-ML are observed near the resonance regime in the low-coupling case, where subtle structural features of the events become more relevant, while in the high-coupling case, the exclusion contours remain almost indistinguishable across the full mass range. The comparison between SR-A (see figure \ref{fig:exclusion_contours}(a))and SR-B see figure \ref{fig:exclusion_contours}(c)) further indicates that imposing the photon recoil mass window significantly alters the exclusion limit, and most of the parameter space is excluded. Importantly, however, at the level of classification accuracy, TDA-ML consistently outperforms ML, underscoring the central motivation of our study: enriching the feature space with topological descriptors leads to more powerful classifiers. The modest improvement in exclusion reach observed here should therefore be viewed as a conservative outcome of testing in a minimal and highly constrained final state. For more complex collider signatures with richer event topologies, the advantages of TDA-ML over conventional ML are expected to become substantially more pronounced \cite{beuria2024intrinsic}, highlighting the potential of topology-inspired methods to enhance the discovery and exclusion capabilities of future high-energy colliders.

\section{Conclusion}
\label{sec:conclusion}
In this work, we have investigated the use of topological data analysis (TDA) in combination with 
machine learning techniques to improve the discrimination between Standard Model backgrounds 
and Beyond Standard Model signals at a muon collider. By representing reconstructed events as 
point clouds in a high-dimensional kinematic space and extracting persistent homology features, 
we have constructed Betti curves that capture the global topological structure of the data. 
Training classifiers at the bag level within the Multiple Instance Learning framework, 
we have demonstrated that the TDA-ML approach achieves consistently higher accuracy than 
conventional ML methods based solely on event-wise kinematic variables. However, our finding suggests that for simpler final states like $\gamma\slashed{E}_T$, the exclusion limits may not tighten compared to the conventional ML approaches.

Looking ahead, these results establish TDA-ML as a promising complementary tool for collider 
phenomenology. Future directions include extending the framework to incorporate alternative signal topologies with complex final states and different filtration strategies to capture more subtle structural differences, and applying the method to other new physics scenarios beyond dark matter production. The integration of TDA with graph-based neural networks or attention mechanisms within MIL architectures may provide 
further gains in discriminative power. As future high-energy colliders push toward increasingly 
challenging parameter spaces, topological descriptors of event ensembles could play a key role 
in strengthening discovery and exclusion potential beyond the reach of traditional 
kinematics-driven analyses.

\bibliography{ref}

\providecommand{\noopsort}[1]{}\providecommand{\singleletter}[1]{#1}%
\begin{thebibliography}{38}%
\makeatletter
\providecommand \@ifxundefined [1]{%
 \@ifx{#1\undefined}
}%
\providecommand \@ifnum [1]{%
 \ifnum #1\expandafter \@firstoftwo
 \else \expandafter \@secondoftwo
 \fi
}%
\providecommand \@ifx [1]{%
 \ifx #1\expandafter \@firstoftwo
 \else \expandafter \@secondoftwo
 \fi
}%
\providecommand \natexlab [1]{#1}%
\providecommand \enquote  [1]{``#1''}%
\providecommand \bibnamefont  [1]{#1}%
\providecommand \bibfnamefont [1]{#1}%
\providecommand \citenamefont [1]{#1}%
\providecommand \href@noop [0]{\@secondoftwo}%
\providecommand \href [0]{\begingroup \@sanitize@url \@href}%
\providecommand \@href[1]{\@@startlink{#1}\@@href}%
\providecommand \@@href[1]{\endgroup#1\@@endlink}%
\providecommand \@sanitize@url [0]{\catcode `\\12\catcode `\$12\catcode `\&12\catcode `\#12\catcode `\^12\catcode `\_12\catcode `\%12\relax}%
\providecommand \@@startlink[1]{}%
\providecommand \@@endlink[0]{}%
\providecommand \url  [0]{\begingroup\@sanitize@url \@url }%
\providecommand \@url [1]{\endgroup\@href {#1}{\urlprefix }}%
\providecommand \urlprefix  [0]{URL }%
\providecommand \Eprint [0]{\href }%
\providecommand \doibase [0]{https://doi.org/}%
\providecommand \selectlanguage [0]{\@gobble}%
\providecommand \bibinfo  [0]{\@secondoftwo}%
\providecommand \bibfield  [0]{\@secondoftwo}%
\providecommand \translation [1]{[#1]}%
\providecommand \BibitemOpen [0]{}%
\providecommand \bibitemStop [0]{}%
\providecommand \bibitemNoStop [0]{.\EOS\space}%
\providecommand \EOS [0]{\spacefactor3000\relax}%
\providecommand \BibitemShut  [1]{\csname bibitem#1\endcsname}%
\let\auto@bib@innerbib\@empty
\bibitem [{\citenamefont {Langacker}(2009)}]{langacker2009physics}%
  \BibitemOpen
  \bibfield  {author} {\bibinfo {author} {\bibfnamefont {P.}~\bibnamefont {Langacker}},\ }\bibfield  {title} {\bibinfo {title} {The physics of heavy z' gauge bosons},\ }\href@noop {} {\bibfield  {journal} {\bibinfo  {journal} {Reviews of Modern Physics}\ }\textbf {\bibinfo {volume} {81}},\ \bibinfo {pages} {1199} (\bibinfo {year} {2009})}\BibitemShut {NoStop}%
\bibitem [{\citenamefont {Robinett}\ and\ \citenamefont {Rosner}(1982)}]{robinett1982prospects}%
  \BibitemOpen
  \bibfield  {author} {\bibinfo {author} {\bibfnamefont {R.~W.}\ \bibnamefont {Robinett}}\ and\ \bibinfo {author} {\bibfnamefont {J.~L.}\ \bibnamefont {Rosner}},\ }\bibfield  {title} {\bibinfo {title} {Prospects for a second neutral vector boson at low mass in so (10)},\ }\href@noop {} {\bibfield  {journal} {\bibinfo  {journal} {Physical Review D}\ }\textbf {\bibinfo {volume} {25}},\ \bibinfo {pages} {3036} (\bibinfo {year} {1982})}\BibitemShut {NoStop}%
\bibitem [{\citenamefont {Hewett}\ and\ \citenamefont {Rizzo}(1989)}]{hewett1989low}%
  \BibitemOpen
  \bibfield  {author} {\bibinfo {author} {\bibfnamefont {J.~L.}\ \bibnamefont {Hewett}}\ and\ \bibinfo {author} {\bibfnamefont {T.~G.}\ \bibnamefont {Rizzo}},\ }\bibfield  {title} {\bibinfo {title} {Low-energy phenomenology of superstring-inspired e6 models},\ }\href@noop {} {\bibfield  {journal} {\bibinfo  {journal} {Physics Reports}\ }\textbf {\bibinfo {volume} {183}},\ \bibinfo {pages} {193} (\bibinfo {year} {1989})}\BibitemShut {NoStop}%
\bibitem [{\citenamefont {Cveti{\v{c}}}\ and\ \citenamefont {Langacker}(1996)}]{cvetivc1996new}%
  \BibitemOpen
  \bibfield  {author} {\bibinfo {author} {\bibfnamefont {M.}~\bibnamefont {Cveti{\v{c}}}}\ and\ \bibinfo {author} {\bibfnamefont {P.}~\bibnamefont {Langacker}},\ }\bibfield  {title} {\bibinfo {title} {New gauge bosons from string models},\ }\href@noop {} {\bibfield  {journal} {\bibinfo  {journal} {Modern Physics Letters A}\ }\textbf {\bibinfo {volume} {11}},\ \bibinfo {pages} {1247} (\bibinfo {year} {1996})}\BibitemShut {NoStop}%
\bibitem [{\citenamefont {Arun}\ \emph {et~al.}(2022)\citenamefont {Arun}, \citenamefont {Chatterjee}, \citenamefont {Mandal}, \citenamefont {Mitra}, \citenamefont {Mukherjee},\ and\ \citenamefont {Nivedita}}]{arun2022search}%
  \BibitemOpen
  \bibfield  {author} {\bibinfo {author} {\bibfnamefont {M.~T.}\ \bibnamefont {Arun}}, \bibinfo {author} {\bibfnamefont {A.}~\bibnamefont {Chatterjee}}, \bibinfo {author} {\bibfnamefont {T.}~\bibnamefont {Mandal}}, \bibinfo {author} {\bibfnamefont {S.}~\bibnamefont {Mitra}}, \bibinfo {author} {\bibfnamefont {A.}~\bibnamefont {Mukherjee}},\ and\ \bibinfo {author} {\bibfnamefont {K.}~\bibnamefont {Nivedita}},\ }\bibfield  {title} {\bibinfo {title} {Search for the z' boson decaying to a right-handed neutrino pair in leptophobic u (1) models},\ }\href@noop {} {\bibfield  {journal} {\bibinfo  {journal} {Physical Review D}\ }\textbf {\bibinfo {volume} {106}},\ \bibinfo {pages} {095035} (\bibinfo {year} {2022})}\BibitemShut {NoStop}%
\bibitem [{\citenamefont {Su}\ \emph {et~al.}(2024)\citenamefont {Su}, \citenamefont {Cai}, \citenamefont {Zeng},\ and\ \citenamefont {Zhang}}]{su2024complex}%
  \BibitemOpen
  \bibfield  {author} {\bibinfo {author} {\bibfnamefont {Y.-H.}\ \bibnamefont {Su}}, \bibinfo {author} {\bibfnamefont {C.}~\bibnamefont {Cai}}, \bibinfo {author} {\bibfnamefont {Y.-P.}\ \bibnamefont {Zeng}},\ and\ \bibinfo {author} {\bibfnamefont {H.-H.}\ \bibnamefont {Zhang}},\ }\bibfield  {title} {\bibinfo {title} {Complex scalar dark matter in a new gauged u (1) symmetry with kinetic and direct mixings},\ }\href@noop {} {\bibfield  {journal} {\bibinfo  {journal} {Physical Review D}\ }\textbf {\bibinfo {volume} {110}},\ \bibinfo {pages} {095014} (\bibinfo {year} {2024})}\BibitemShut {NoStop}%
\bibitem [{\citenamefont {Barik}\ \emph {et~al.}(2025)\citenamefont {Barik}, \citenamefont {Rai},\ and\ \citenamefont {Srivastava}}]{barik2025discovering}%
  \BibitemOpen
  \bibfield  {author} {\bibinfo {author} {\bibfnamefont {A.~K.}\ \bibnamefont {Barik}}, \bibinfo {author} {\bibfnamefont {S.~K.}\ \bibnamefont {Rai}},\ and\ \bibinfo {author} {\bibfnamefont {A.}~\bibnamefont {Srivastava}},\ }\bibfield  {title} {\bibinfo {title} {Discovering an invisible z' at the muon collider},\ }\href@noop {} {\bibfield  {journal} {\bibinfo  {journal} {Physics Letters B}\ ,\ \bibinfo {pages} {139533}} (\bibinfo {year} {2025})}\BibitemShut {NoStop}%
\bibitem [{\citenamefont {Das}\ \emph {et~al.}(2025)\citenamefont {Das}, \citenamefont {Das},\ and\ \citenamefont {Okada}}]{das2025testing}%
  \BibitemOpen
  \bibfield  {author} {\bibinfo {author} {\bibfnamefont {A.}~\bibnamefont {Das}}, \bibinfo {author} {\bibfnamefont {P.}~\bibnamefont {Das}},\ and\ \bibinfo {author} {\bibfnamefont {N.}~\bibnamefont {Okada}},\ }\bibfield  {title} {\bibinfo {title} {Testing neutrino mass hierarchy under type-ii seesaw scenario in u (1) x from colliders},\ }\href@noop {} {\bibfield  {journal} {\bibinfo  {journal} {Physics Letters B}\ ,\ \bibinfo {pages} {139475}} (\bibinfo {year} {2025})}\BibitemShut {NoStop}%
\bibitem [{\citenamefont {Alves}\ \emph {et~al.}(2015)\citenamefont {Alves}, \citenamefont {Berlin}, \citenamefont {Profumo},\ and\ \citenamefont {Queiroz}}]{alves2015dark}%
  \BibitemOpen
  \bibfield  {author} {\bibinfo {author} {\bibfnamefont {A.}~\bibnamefont {Alves}}, \bibinfo {author} {\bibfnamefont {A.}~\bibnamefont {Berlin}}, \bibinfo {author} {\bibfnamefont {S.}~\bibnamefont {Profumo}},\ and\ \bibinfo {author} {\bibfnamefont {F.~S.}\ \bibnamefont {Queiroz}},\ }\bibfield  {title} {\bibinfo {title} {Dark matter complementarity and the z' portal},\ }\href@noop {} {\bibfield  {journal} {\bibinfo  {journal} {Physical Review D}\ }\textbf {\bibinfo {volume} {92}},\ \bibinfo {pages} {083004} (\bibinfo {year} {2015})}\BibitemShut {NoStop}%
\bibitem [{\citenamefont {{\.I}nan}\ and\ \citenamefont {Kisselev}(2022)}]{inan2022search}%
  \BibitemOpen
  \bibfield  {author} {\bibinfo {author} {\bibfnamefont {S.}~\bibnamefont {{\.I}nan}}\ and\ \bibinfo {author} {\bibfnamefont {A.}~\bibnamefont {Kisselev}},\ }\bibfield  {title} {\bibinfo {title} {Search for invisible dark photon in $\gamma$ e scattering at future lepton colliders},\ }\href@noop {} {\bibfield  {journal} {\bibinfo  {journal} {The European Physical Journal C}\ }\textbf {\bibinfo {volume} {82}},\ \bibinfo {pages} {592} (\bibinfo {year} {2022})}\BibitemShut {NoStop}%
\bibitem [{\citenamefont {Wang}\ \emph {et~al.}(2025)\citenamefont {Wang}, \citenamefont {Han}, \citenamefont {Huang}, \citenamefont {Jin},\ and\ \citenamefont {Li}}]{wang2025prospects}%
  \BibitemOpen
  \bibfield  {author} {\bibinfo {author} {\bibfnamefont {Z.-W.}\ \bibnamefont {Wang}}, \bibinfo {author} {\bibfnamefont {Z.-L.}\ \bibnamefont {Han}}, \bibinfo {author} {\bibfnamefont {F.}~\bibnamefont {Huang}}, \bibinfo {author} {\bibfnamefont {Y.}~\bibnamefont {Jin}},\ and\ \bibinfo {author} {\bibfnamefont {H.}~\bibnamefont {Li}},\ }\bibfield  {title} {\bibinfo {title} {Prospects of z' portal dark matter in u (1) l $\mu$-l $\tau$},\ }\href@noop {} {\bibfield  {journal} {\bibinfo  {journal} {Physical Review D}\ }\textbf {\bibinfo {volume} {111}},\ \bibinfo {pages} {095017} (\bibinfo {year} {2025})}\BibitemShut {NoStop}%
\bibitem [{\citenamefont {Feng}\ and\ \citenamefont {Zhang}(2025)}]{feng2025dark}%
  \BibitemOpen
  \bibfield  {author} {\bibinfo {author} {\bibfnamefont {W.-Z.}\ \bibnamefont {Feng}}\ and\ \bibinfo {author} {\bibfnamefont {Z.-H.}\ \bibnamefont {Zhang}},\ }\bibfield  {title} {\bibinfo {title} {Dark matter generation from a dark sector},\ }\href@noop {} {\bibfield  {journal} {\bibinfo  {journal} {Physical Review D}\ }\textbf {\bibinfo {volume} {112}},\ \bibinfo {pages} {035004} (\bibinfo {year} {2025})}\BibitemShut {NoStop}%
\bibitem [{\citenamefont {Aad}\ \emph {et~al.}(2019)\citenamefont {Aad} \emph {et~al.}}]{ATLAS:2019erb}%
  \BibitemOpen
  \bibfield  {author} {\bibinfo {author} {\bibfnamefont {G.}~\bibnamefont {Aad}} \emph {et~al.} (\bibinfo {collaboration} {ATLAS}),\ }\bibfield  {title} {\bibinfo {title} {{Search for high-mass dilepton resonances using 139 fb$^{-1}$ of $pp$ collision data collected at $\sqrt{s}=$13 TeV with the ATLAS detector}},\ }\href {https://doi.org/10.1016/j.physletb.2019.07.016} {\bibfield  {journal} {\bibinfo  {journal} {Phys. Lett. B}\ }\textbf {\bibinfo {volume} {796}},\ \bibinfo {pages} {68} (\bibinfo {year} {2019})},\ \Eprint {https://arxiv.org/abs/1903.06248} {arXiv:1903.06248 [hep-ex]} \BibitemShut {NoStop}%
\bibitem [{\citenamefont {Sirunyan}\ \emph {et~al.}(2018)\citenamefont {Sirunyan} \emph {et~al.}}]{CMS:2018ipm}%
  \BibitemOpen
  \bibfield  {author} {\bibinfo {author} {\bibfnamefont {A.~M.}\ \bibnamefont {Sirunyan}} \emph {et~al.} (\bibinfo {collaboration} {CMS}),\ }\bibfield  {title} {\bibinfo {title} {{Search for high-mass resonances in dilepton final states in proton-proton collisions at $\sqrt{s}=$ 13 TeV}},\ }\href {https://doi.org/10.1007/JHEP06(2018)120} {\bibfield  {journal} {\bibinfo  {journal} {JHEP}\ }\textbf {\bibinfo {volume} {06}},\ \bibinfo {pages} {120}},\ \Eprint {https://arxiv.org/abs/1803.06292} {arXiv:1803.06292 [hep-ex]} \BibitemShut {NoStop}%
\bibitem [{\citenamefont {Sirunyan}\ \emph {et~al.}(2020)\citenamefont {Sirunyan}, \citenamefont {Tumasyan}, \citenamefont {Adam}, \citenamefont {Ambrogi}, \citenamefont {Bergauer}, \citenamefont {Brandstetter}, \citenamefont {Dragicevic}, \citenamefont {Er{\"o}}, \citenamefont {Del~Valle}, \citenamefont {Flechl} \emph {et~al.}}]{sirunyan2020multi}%
  \BibitemOpen
  \bibfield  {author} {\bibinfo {author} {\bibfnamefont {A.~M.}\ \bibnamefont {Sirunyan}}, \bibinfo {author} {\bibfnamefont {A.}~\bibnamefont {Tumasyan}}, \bibinfo {author} {\bibfnamefont {W.}~\bibnamefont {Adam}}, \bibinfo {author} {\bibfnamefont {F.}~\bibnamefont {Ambrogi}}, \bibinfo {author} {\bibfnamefont {T.}~\bibnamefont {Bergauer}}, \bibinfo {author} {\bibfnamefont {J.}~\bibnamefont {Brandstetter}}, \bibinfo {author} {\bibfnamefont {M.}~\bibnamefont {Dragicevic}}, \bibinfo {author} {\bibfnamefont {J.}~\bibnamefont {Er{\"o}}}, \bibinfo {author} {\bibfnamefont {A.~E.}\ \bibnamefont {Del~Valle}}, \bibinfo {author} {\bibfnamefont {M.}~\bibnamefont {Flechl}}, \emph {et~al.},\ }\bibfield  {title} {\bibinfo {title} {A multi-dimensional search for new heavy resonances decaying to boosted ww, wz, or zz boson pairs in the dijet final state at 13 te},\ }\href@noop {} {\bibfield  {journal} {\bibinfo  {journal} {The European Physical Journal C}\ }\textbf {\bibinfo {volume} {80}},\ \bibinfo {pages} {237} (\bibinfo
  {year} {2020})}\BibitemShut {NoStop}%
\bibitem [{\citenamefont {Alves}\ \emph {et~al.}(2014)\citenamefont {Alves}, \citenamefont {Profumo},\ and\ \citenamefont {Queiroz}}]{alves2014dark}%
  \BibitemOpen
  \bibfield  {author} {\bibinfo {author} {\bibfnamefont {A.}~\bibnamefont {Alves}}, \bibinfo {author} {\bibfnamefont {S.}~\bibnamefont {Profumo}},\ and\ \bibinfo {author} {\bibfnamefont {F.~S.}\ \bibnamefont {Queiroz}},\ }\bibfield  {title} {\bibinfo {title} {The dark z' portal: direct, indirect and collider searches},\ }\href@noop {} {\bibfield  {journal} {\bibinfo  {journal} {Journal of High Energy Physics}\ }\textbf {\bibinfo {volume} {2014}},\ \bibinfo {pages} {1} (\bibinfo {year} {2014})}\BibitemShut {NoStop}%
\bibitem [{\citenamefont {Baer}\ \emph {et~al.}(2013)\citenamefont {Baer}, \citenamefont {Barklow}, \citenamefont {Fujii}, \citenamefont {Gao}, \citenamefont {Hoang}, \citenamefont {Kanemura}, \citenamefont {List}, \citenamefont {Logan}, \citenamefont {Nomerotski}, \citenamefont {Perelstein} \emph {et~al.}}]{baer2013international}%
  \BibitemOpen
  \bibfield  {author} {\bibinfo {author} {\bibfnamefont {H.}~\bibnamefont {Baer}}, \bibinfo {author} {\bibfnamefont {T.}~\bibnamefont {Barklow}}, \bibinfo {author} {\bibfnamefont {K.}~\bibnamefont {Fujii}}, \bibinfo {author} {\bibfnamefont {Y.}~\bibnamefont {Gao}}, \bibinfo {author} {\bibfnamefont {A.}~\bibnamefont {Hoang}}, \bibinfo {author} {\bibfnamefont {S.}~\bibnamefont {Kanemura}}, \bibinfo {author} {\bibfnamefont {J.}~\bibnamefont {List}}, \bibinfo {author} {\bibfnamefont {H.~E.}\ \bibnamefont {Logan}}, \bibinfo {author} {\bibfnamefont {A.}~\bibnamefont {Nomerotski}}, \bibinfo {author} {\bibfnamefont {M.}~\bibnamefont {Perelstein}}, \emph {et~al.},\ }\bibfield  {title} {\bibinfo {title} {The international linear collider technical design report-volume 2: physics},\ }\href@noop {} {\bibfield  {journal} {\bibinfo  {journal} {arXiv preprint arXiv:1306.6352}\ } (\bibinfo {year} {2013})}\BibitemShut {NoStop}%
\bibitem [{\citenamefont {Abramowicz}\ \emph {et~al.}(2017)\citenamefont {Abramowicz}, \citenamefont {Abusleme}, \citenamefont {Afanaciev}, \citenamefont {Alipour~Tehrani}, \citenamefont {Bal{\'a}zs}, \citenamefont {Benhammou}, \citenamefont {Benoit}, \citenamefont {Bilki}, \citenamefont {Blaising}, \citenamefont {Boland} \emph {et~al.}}]{abramowicz2017higgs}%
  \BibitemOpen
  \bibfield  {author} {\bibinfo {author} {\bibfnamefont {H.}~\bibnamefont {Abramowicz}}, \bibinfo {author} {\bibfnamefont {A.}~\bibnamefont {Abusleme}}, \bibinfo {author} {\bibfnamefont {K.}~\bibnamefont {Afanaciev}}, \bibinfo {author} {\bibfnamefont {N.}~\bibnamefont {Alipour~Tehrani}}, \bibinfo {author} {\bibfnamefont {C.}~\bibnamefont {Bal{\'a}zs}}, \bibinfo {author} {\bibfnamefont {Y.}~\bibnamefont {Benhammou}}, \bibinfo {author} {\bibfnamefont {M.}~\bibnamefont {Benoit}}, \bibinfo {author} {\bibfnamefont {B.}~\bibnamefont {Bilki}}, \bibinfo {author} {\bibfnamefont {J.-J.}\ \bibnamefont {Blaising}}, \bibinfo {author} {\bibfnamefont {M.}~\bibnamefont {Boland}}, \emph {et~al.},\ }\bibfield  {title} {\bibinfo {title} {Higgs physics at the clic electron--positron linear collider},\ }\href@noop {} {\bibfield  {journal} {\bibinfo  {journal} {The European Physical Journal C}\ }\textbf {\bibinfo {volume} {77}},\ \bibinfo {pages} {1} (\bibinfo {year} {2017})}\BibitemShut {NoStop}%
\bibitem [{\citenamefont {Abada}\ \emph {et~al.}(2019)\citenamefont {Abada}, \citenamefont {Abbrescia}, \citenamefont {AbdusSalam}, \citenamefont {Abdyukhanov}, \citenamefont {Abelleira~Fernandez}, \citenamefont {Abramov}, \citenamefont {Aburaia}, \citenamefont {Acar}, \citenamefont {Adzic}, \citenamefont {Agrawal} \emph {et~al.}}]{abada2019fcc}%
  \BibitemOpen
  \bibfield  {author} {\bibinfo {author} {\bibfnamefont {A.}~\bibnamefont {Abada}}, \bibinfo {author} {\bibfnamefont {M.}~\bibnamefont {Abbrescia}}, \bibinfo {author} {\bibfnamefont {S.}~\bibnamefont {AbdusSalam}}, \bibinfo {author} {\bibfnamefont {I.}~\bibnamefont {Abdyukhanov}}, \bibinfo {author} {\bibfnamefont {J.}~\bibnamefont {Abelleira~Fernandez}}, \bibinfo {author} {\bibfnamefont {A.}~\bibnamefont {Abramov}}, \bibinfo {author} {\bibfnamefont {M.}~\bibnamefont {Aburaia}}, \bibinfo {author} {\bibfnamefont {A.}~\bibnamefont {Acar}}, \bibinfo {author} {\bibfnamefont {P.}~\bibnamefont {Adzic}}, \bibinfo {author} {\bibfnamefont {P.}~\bibnamefont {Agrawal}}, \emph {et~al.},\ }\bibfield  {title} {\bibinfo {title} {Fcc-ee: The lepton collider: Future circular collider conceptual design report volume 2},\ }\href@noop {} {\bibfield  {journal} {\bibinfo  {journal} {The European Physical Journal Special Topics}\ }\textbf {\bibinfo {volume} {228}},\ \bibinfo {pages} {261} (\bibinfo {year} {2019})}\BibitemShut
  {NoStop}%
\bibitem [{\citenamefont {Agapov}\ \emph {et~al.}(2022)\citenamefont {Agapov}, \citenamefont {Benedikt}, \citenamefont {Blondel}, \citenamefont {Boscolo}, \citenamefont {Brunner}, \citenamefont {Llatas}, \citenamefont {Charles}, \citenamefont {Denisov}, \citenamefont {Fischer}, \citenamefont {Gianfelice-Wendt} \emph {et~al.}}]{agapov2022future}%
  \BibitemOpen
  \bibfield  {author} {\bibinfo {author} {\bibfnamefont {I.}~\bibnamefont {Agapov}}, \bibinfo {author} {\bibfnamefont {M.}~\bibnamefont {Benedikt}}, \bibinfo {author} {\bibfnamefont {A.}~\bibnamefont {Blondel}}, \bibinfo {author} {\bibfnamefont {M.}~\bibnamefont {Boscolo}}, \bibinfo {author} {\bibfnamefont {O.}~\bibnamefont {Brunner}}, \bibinfo {author} {\bibfnamefont {M.~C.}\ \bibnamefont {Llatas}}, \bibinfo {author} {\bibfnamefont {T.}~\bibnamefont {Charles}}, \bibinfo {author} {\bibfnamefont {D.}~\bibnamefont {Denisov}}, \bibinfo {author} {\bibfnamefont {W.}~\bibnamefont {Fischer}}, \bibinfo {author} {\bibfnamefont {E.}~\bibnamefont {Gianfelice-Wendt}}, \emph {et~al.},\ }\bibfield  {title} {\bibinfo {title} {Future circular lepton collider fcc-ee: Overview and status},\ }\href@noop {} {\bibfield  {journal} {\bibinfo  {journal} {arXiv preprint arXiv:2203.08310}\ } (\bibinfo {year} {2022})}\BibitemShut {NoStop}%
\bibitem [{\citenamefont {Delahaye}\ \emph {et~al.}(2019)\citenamefont {Delahaye}, \citenamefont {Diemoz}, \citenamefont {Long}, \citenamefont {Mansouli{\'e}}, \citenamefont {Pastrone}, \citenamefont {Rivkin}, \citenamefont {Schulte}, \citenamefont {Skrinsky},\ and\ \citenamefont {Wulzer}}]{delahaye2019muon}%
  \BibitemOpen
  \bibfield  {author} {\bibinfo {author} {\bibfnamefont {J.~P.}\ \bibnamefont {Delahaye}}, \bibinfo {author} {\bibfnamefont {M.}~\bibnamefont {Diemoz}}, \bibinfo {author} {\bibfnamefont {K.}~\bibnamefont {Long}}, \bibinfo {author} {\bibfnamefont {B.}~\bibnamefont {Mansouli{\'e}}}, \bibinfo {author} {\bibfnamefont {N.}~\bibnamefont {Pastrone}}, \bibinfo {author} {\bibfnamefont {L.}~\bibnamefont {Rivkin}}, \bibinfo {author} {\bibfnamefont {D.}~\bibnamefont {Schulte}}, \bibinfo {author} {\bibfnamefont {A.}~\bibnamefont {Skrinsky}},\ and\ \bibinfo {author} {\bibfnamefont {A.}~\bibnamefont {Wulzer}},\ }\bibfield  {title} {\bibinfo {title} {Muon colliders},\ }\href@noop {} {\bibfield  {journal} {\bibinfo  {journal} {arXiv preprint arXiv:1901.06150}\ } (\bibinfo {year} {2019})}\BibitemShut {NoStop}%
\bibitem [{\citenamefont {Long}\ \emph {et~al.}(2021)\citenamefont {Long}, \citenamefont {Lucchesi}, \citenamefont {Palmer}, \citenamefont {Pastrone}, \citenamefont {Schulte},\ and\ \citenamefont {Shiltsev}}]{long2021muon}%
  \BibitemOpen
  \bibfield  {author} {\bibinfo {author} {\bibfnamefont {K.~R.}\ \bibnamefont {Long}}, \bibinfo {author} {\bibfnamefont {D.}~\bibnamefont {Lucchesi}}, \bibinfo {author} {\bibfnamefont {M.~A.}\ \bibnamefont {Palmer}}, \bibinfo {author} {\bibfnamefont {N.}~\bibnamefont {Pastrone}}, \bibinfo {author} {\bibfnamefont {D.}~\bibnamefont {Schulte}},\ and\ \bibinfo {author} {\bibfnamefont {V.}~\bibnamefont {Shiltsev}},\ }\bibfield  {title} {\bibinfo {title} {Muon colliders to expand frontiers of particle physics},\ }\href@noop {} {\bibfield  {journal} {\bibinfo  {journal} {Nature Physics}\ }\textbf {\bibinfo {volume} {17}},\ \bibinfo {pages} {289} (\bibinfo {year} {2021})}\BibitemShut {NoStop}%
\bibitem [{\citenamefont {Accettura}\ \emph {et~al.}(2023)\citenamefont {Accettura}, \citenamefont {Adams}, \citenamefont {Agarwal}, \citenamefont {Ahdida}, \citenamefont {Aim{\`e}}, \citenamefont {Amapane}, \citenamefont {Amorim}, \citenamefont {Andreetto}, \citenamefont {Anulli}, \citenamefont {Appleby} \emph {et~al.}}]{accettura2023towards}%
  \BibitemOpen
  \bibfield  {author} {\bibinfo {author} {\bibfnamefont {C.}~\bibnamefont {Accettura}}, \bibinfo {author} {\bibfnamefont {D.}~\bibnamefont {Adams}}, \bibinfo {author} {\bibfnamefont {R.}~\bibnamefont {Agarwal}}, \bibinfo {author} {\bibfnamefont {C.}~\bibnamefont {Ahdida}}, \bibinfo {author} {\bibfnamefont {C.}~\bibnamefont {Aim{\`e}}}, \bibinfo {author} {\bibfnamefont {N.}~\bibnamefont {Amapane}}, \bibinfo {author} {\bibfnamefont {D.}~\bibnamefont {Amorim}}, \bibinfo {author} {\bibfnamefont {P.}~\bibnamefont {Andreetto}}, \bibinfo {author} {\bibfnamefont {F.}~\bibnamefont {Anulli}}, \bibinfo {author} {\bibfnamefont {R.}~\bibnamefont {Appleby}}, \emph {et~al.},\ }\bibfield  {title} {\bibinfo {title} {Towards a muon collider},\ }\href@noop {} {\bibfield  {journal} {\bibinfo  {journal} {The European Physical Journal C}\ }\textbf {\bibinfo {volume} {83}},\ \bibinfo {pages} {864} (\bibinfo {year} {2023})}\BibitemShut {NoStop}%
\bibitem [{\citenamefont {Binner}\ \emph {et~al.}(1999)\citenamefont {Binner}, \citenamefont {Kuhn},\ and\ \citenamefont {Melnikov}}]{Binner:1999bt}%
  \BibitemOpen
  \bibfield  {author} {\bibinfo {author} {\bibfnamefont {S.}~\bibnamefont {Binner}}, \bibinfo {author} {\bibfnamefont {J.~H.}\ \bibnamefont {Kuhn}},\ and\ \bibinfo {author} {\bibfnamefont {K.}~\bibnamefont {Melnikov}},\ }\bibfield  {title} {\bibinfo {title} {{Measuring sigma(e+ e- ---{\ensuremath{>}} hadrons) using tagged photon}},\ }\href {https://doi.org/10.1016/S0370-2693(99)00658-9} {\bibfield  {journal} {\bibinfo  {journal} {Phys. Lett. B}\ }\textbf {\bibinfo {volume} {459}},\ \bibinfo {pages} {279} (\bibinfo {year} {1999})},\ \Eprint {https://arxiv.org/abs/hep-ph/9902399} {arXiv:hep-ph/9902399} \BibitemShut {NoStop}%
\bibitem [{\citenamefont {Radovic}\ \emph {et~al.}(2018)\citenamefont {Radovic}, \citenamefont {Williams}, \citenamefont {Rousseau}, \citenamefont {Kagan}, \citenamefont {Bonacorsi}, \citenamefont {Himmel}, \citenamefont {Aurisano}, \citenamefont {Terao},\ and\ \citenamefont {Wongjirad}}]{radovic2018machine}%
  \BibitemOpen
  \bibfield  {author} {\bibinfo {author} {\bibfnamefont {A.}~\bibnamefont {Radovic}}, \bibinfo {author} {\bibfnamefont {M.}~\bibnamefont {Williams}}, \bibinfo {author} {\bibfnamefont {D.}~\bibnamefont {Rousseau}}, \bibinfo {author} {\bibfnamefont {M.}~\bibnamefont {Kagan}}, \bibinfo {author} {\bibfnamefont {D.}~\bibnamefont {Bonacorsi}}, \bibinfo {author} {\bibfnamefont {A.}~\bibnamefont {Himmel}}, \bibinfo {author} {\bibfnamefont {A.}~\bibnamefont {Aurisano}}, \bibinfo {author} {\bibfnamefont {K.}~\bibnamefont {Terao}},\ and\ \bibinfo {author} {\bibfnamefont {T.}~\bibnamefont {Wongjirad}},\ }\bibfield  {title} {\bibinfo {title} {Machine learning at the energy and intensity frontiers of particle physics},\ }\href@noop {} {\bibfield  {journal} {\bibinfo  {journal} {Nature}\ }\textbf {\bibinfo {volume} {560}},\ \bibinfo {pages} {41} (\bibinfo {year} {2018})}\BibitemShut {NoStop}%
\bibitem [{\citenamefont {Schwartz}(2021)}]{schwartz2021modern}%
  \BibitemOpen
  \bibfield  {author} {\bibinfo {author} {\bibfnamefont {M.~D.}\ \bibnamefont {Schwartz}},\ }\bibfield  {title} {\bibinfo {title} {Modern machine learning and particle physics},\ }\href@noop {} {\bibfield  {journal} {\bibinfo  {journal} {arXiv preprint arXiv:2103.12226}\ } (\bibinfo {year} {2021})}\BibitemShut {NoStop}%
\bibitem [{\citenamefont {Karagiorgi}\ \emph {et~al.}(2022)\citenamefont {Karagiorgi}, \citenamefont {Kasieczka}, \citenamefont {Kravitz}, \citenamefont {Nachman},\ and\ \citenamefont {Shih}}]{karagiorgi2022machine}%
  \BibitemOpen
  \bibfield  {author} {\bibinfo {author} {\bibfnamefont {G.}~\bibnamefont {Karagiorgi}}, \bibinfo {author} {\bibfnamefont {G.}~\bibnamefont {Kasieczka}}, \bibinfo {author} {\bibfnamefont {S.}~\bibnamefont {Kravitz}}, \bibinfo {author} {\bibfnamefont {B.}~\bibnamefont {Nachman}},\ and\ \bibinfo {author} {\bibfnamefont {D.}~\bibnamefont {Shih}},\ }\bibfield  {title} {\bibinfo {title} {Machine learning in the search for new fundamental physics},\ }\href@noop {} {\bibfield  {journal} {\bibinfo  {journal} {Nature Reviews Physics}\ }\textbf {\bibinfo {volume} {4}},\ \bibinfo {pages} {399} (\bibinfo {year} {2022})}\BibitemShut {NoStop}%
\bibitem [{\citenamefont {Chazal}\ and\ \citenamefont {Michel}(2021)}]{chazal2021introduction}%
  \BibitemOpen
  \bibfield  {author} {\bibinfo {author} {\bibfnamefont {F.}~\bibnamefont {Chazal}}\ and\ \bibinfo {author} {\bibfnamefont {B.}~\bibnamefont {Michel}},\ }\bibfield  {title} {\bibinfo {title} {An introduction to topological data analysis: fundamental and practical aspects for data scientists},\ }\href@noop {} {\bibfield  {journal} {\bibinfo  {journal} {Frontiers in artificial intelligence}\ }\textbf {\bibinfo {volume} {4}},\ \bibinfo {pages} {108} (\bibinfo {year} {2021})}\BibitemShut {NoStop}%
\bibitem [{\citenamefont {Beuria}(2023)}]{beuria2023persistent}%
  \BibitemOpen
  \bibfield  {author} {\bibinfo {author} {\bibfnamefont {J.}~\bibnamefont {Beuria}},\ }\bibfield  {title} {\bibinfo {title} {Persistent homology of collider observations: When (w) hole matters},\ }\href@noop {} {\bibfield  {journal} {\bibinfo  {journal} {Physics Letters B}\ }\textbf {\bibinfo {volume} {846}},\ \bibinfo {pages} {138188} (\bibinfo {year} {2023})}\BibitemShut {NoStop}%
\bibitem [{\citenamefont {Beuria}(2024)}]{beuria2024intrinsic}%
  \BibitemOpen
  \bibfield  {author} {\bibinfo {author} {\bibfnamefont {J.}~\bibnamefont {Beuria}},\ }\bibfield  {title} {\bibinfo {title} {Intrinsic geometry of collider observations and forman ricci curvature},\ }\href@noop {} {\bibfield  {journal} {\bibinfo  {journal} {Physical Review D}\ }\textbf {\bibinfo {volume} {110}},\ \bibinfo {pages} {035023} (\bibinfo {year} {2024})}\BibitemShut {NoStop}%
\bibitem [{\citenamefont {Carbonneau}\ \emph {et~al.}(2018)\citenamefont {Carbonneau}, \citenamefont {Cheplygina}, \citenamefont {Granger},\ and\ \citenamefont {Gagnon}}]{carbonneau2018multiple}%
  \BibitemOpen
  \bibfield  {author} {\bibinfo {author} {\bibfnamefont {M.-A.}\ \bibnamefont {Carbonneau}}, \bibinfo {author} {\bibfnamefont {V.}~\bibnamefont {Cheplygina}}, \bibinfo {author} {\bibfnamefont {E.}~\bibnamefont {Granger}},\ and\ \bibinfo {author} {\bibfnamefont {G.}~\bibnamefont {Gagnon}},\ }\bibfield  {title} {\bibinfo {title} {Multiple instance learning: A survey of problem characteristics and applications},\ }\href@noop {} {\bibfield  {journal} {\bibinfo  {journal} {Pattern recognition}\ }\textbf {\bibinfo {volume} {77}},\ \bibinfo {pages} {329} (\bibinfo {year} {2018})}\BibitemShut {NoStop}%
\bibitem [{\citenamefont {Carlsson}(2009)}]{carlsson2009topology}%
  \BibitemOpen
  \bibfield  {author} {\bibinfo {author} {\bibfnamefont {G.}~\bibnamefont {Carlsson}},\ }\bibfield  {title} {\bibinfo {title} {Topology and data},\ }\href@noop {} {\bibfield  {journal} {\bibinfo  {journal} {Bulletin of the American Mathematical Society}\ }\textbf {\bibinfo {volume} {46}},\ \bibinfo {pages} {255} (\bibinfo {year} {2009})}\BibitemShut {NoStop}%
\bibitem [{\citenamefont {Murugan}\ and\ \citenamefont {Robertson}(2019)}]{murugan2019introduction}%
  \BibitemOpen
  \bibfield  {author} {\bibinfo {author} {\bibfnamefont {J.}~\bibnamefont {Murugan}}\ and\ \bibinfo {author} {\bibfnamefont {D.}~\bibnamefont {Robertson}},\ }\bibfield  {title} {\bibinfo {title} {An introduction to topological data analysis for physicists: From lgm to frbs},\ }\href@noop {} {\bibfield  {journal} {\bibinfo  {journal} {arXiv preprint arXiv:1904.11044}\ } (\bibinfo {year} {2019})}\BibitemShut {NoStop}%
\bibitem [{\citenamefont {Carlsson}(2020)}]{carlsson2020topological}%
  \BibitemOpen
  \bibfield  {author} {\bibinfo {author} {\bibfnamefont {G.}~\bibnamefont {Carlsson}},\ }\bibfield  {title} {\bibinfo {title} {Topological methods for data modelling},\ }\href@noop {} {\bibfield  {journal} {\bibinfo  {journal} {Nature Reviews Physics}\ }\textbf {\bibinfo {volume} {2}},\ \bibinfo {pages} {697} (\bibinfo {year} {2020})}\BibitemShut {NoStop}%
\bibitem [{\citenamefont {Alwall}\ \emph {et~al.}(2014)\citenamefont {Alwall}, \citenamefont {Frederix}, \citenamefont {Frixione}, \citenamefont {Hirschi}, \citenamefont {Maltoni}, \citenamefont {Mattelaer}, \citenamefont {Shao}, \citenamefont {Stelzer}, \citenamefont {Torrielli},\ and\ \citenamefont {Zaro}}]{Alwall:2014hca}%
  \BibitemOpen
  \bibfield  {author} {\bibinfo {author} {\bibfnamefont {J.}~\bibnamefont {Alwall}}, \bibinfo {author} {\bibfnamefont {R.}~\bibnamefont {Frederix}}, \bibinfo {author} {\bibfnamefont {S.}~\bibnamefont {Frixione}}, \bibinfo {author} {\bibfnamefont {V.}~\bibnamefont {Hirschi}}, \bibinfo {author} {\bibfnamefont {F.}~\bibnamefont {Maltoni}}, \bibinfo {author} {\bibfnamefont {O.}~\bibnamefont {Mattelaer}}, \bibinfo {author} {\bibfnamefont {H.~S.}\ \bibnamefont {Shao}}, \bibinfo {author} {\bibfnamefont {T.}~\bibnamefont {Stelzer}}, \bibinfo {author} {\bibfnamefont {P.}~\bibnamefont {Torrielli}},\ and\ \bibinfo {author} {\bibfnamefont {M.}~\bibnamefont {Zaro}},\ }\bibfield  {title} {\bibinfo {title} {{The automated computation of tree-level and next-to-leading order differential cross sections, and their matching to parton shower simulations}},\ }\href {https://doi.org/10.1007/JHEP07(2014)079} {\bibfield  {journal} {\bibinfo  {journal} {JHEP}\ }\textbf {\bibinfo {volume} {07}},\ \bibinfo {pages} {079}},\ \Eprint
  {https://arxiv.org/abs/1405.0301} {arXiv:1405.0301 [hep-ph]} \BibitemShut {NoStop}%
\bibitem [{\citenamefont {Frederix}\ \emph {et~al.}(2018)\citenamefont {Frederix}, \citenamefont {Frixione}, \citenamefont {Hirschi}, \citenamefont {Pagani}, \citenamefont {Shao},\ and\ \citenamefont {Zaro}}]{Frederix:2018nkq}%
  \BibitemOpen
  \bibfield  {author} {\bibinfo {author} {\bibfnamefont {R.}~\bibnamefont {Frederix}}, \bibinfo {author} {\bibfnamefont {S.}~\bibnamefont {Frixione}}, \bibinfo {author} {\bibfnamefont {V.}~\bibnamefont {Hirschi}}, \bibinfo {author} {\bibfnamefont {D.}~\bibnamefont {Pagani}}, \bibinfo {author} {\bibfnamefont {H.~S.}\ \bibnamefont {Shao}},\ and\ \bibinfo {author} {\bibfnamefont {M.}~\bibnamefont {Zaro}},\ }\bibfield  {title} {\bibinfo {title} {{The automation of next-to-leading order electroweak calculations}},\ }\href {https://doi.org/10.1007/JHEP11(2021)085} {\bibfield  {journal} {\bibinfo  {journal} {JHEP}\ }\textbf {\bibinfo {volume} {07}},\ \bibinfo {pages} {185}},\ \bibinfo {note} {[Erratum: JHEP 11, 085 (2021)]},\ \Eprint {https://arxiv.org/abs/1804.10017} {arXiv:1804.10017 [hep-ph]} \BibitemShut {NoStop}%
\bibitem [{\citenamefont {Bierlich}\ \emph {et~al.}(2022)\citenamefont {Bierlich}, \citenamefont {Chakraborty}, \citenamefont {Desai}, \citenamefont {Gellersen}, \citenamefont {Helenius}, \citenamefont {Ilten}, \citenamefont {L{\"o}nnblad}, \citenamefont {Mrenna}, \citenamefont {Prestel}, \citenamefont {Preuss} \emph {et~al.}}]{bierlich2022comprehensive}%
  \BibitemOpen
  \bibfield  {author} {\bibinfo {author} {\bibfnamefont {C.}~\bibnamefont {Bierlich}}, \bibinfo {author} {\bibfnamefont {S.}~\bibnamefont {Chakraborty}}, \bibinfo {author} {\bibfnamefont {N.}~\bibnamefont {Desai}}, \bibinfo {author} {\bibfnamefont {L.}~\bibnamefont {Gellersen}}, \bibinfo {author} {\bibfnamefont {I.}~\bibnamefont {Helenius}}, \bibinfo {author} {\bibfnamefont {P.}~\bibnamefont {Ilten}}, \bibinfo {author} {\bibfnamefont {L.}~\bibnamefont {L{\"o}nnblad}}, \bibinfo {author} {\bibfnamefont {S.}~\bibnamefont {Mrenna}}, \bibinfo {author} {\bibfnamefont {S.}~\bibnamefont {Prestel}}, \bibinfo {author} {\bibfnamefont {C.~T.}\ \bibnamefont {Preuss}}, \emph {et~al.},\ }\bibfield  {title} {\bibinfo {title} {A comprehensive guide to the physics and usage of pythia 8.3},\ }\href@noop {} {\bibfield  {journal} {\bibinfo  {journal} {SciPost Physics Codebases}\ ,\ \bibinfo {pages} {008}} (\bibinfo {year} {2022})}\BibitemShut {NoStop}%
\bibitem [{\citenamefont {de~Favereau}\ \emph {et~al.}(2014)\citenamefont {de~Favereau}, \citenamefont {Delaere}, \citenamefont {Demin}, \citenamefont {Giammanco}, \citenamefont {Lema\^\i{}tre}, \citenamefont {Mertens},\ and\ \citenamefont {Selvaggi}}]{deFavereau:2013fsa}%
  \BibitemOpen
  \bibfield  {author} {\bibinfo {author} {\bibfnamefont {J.}~\bibnamefont {de~Favereau}}, \bibinfo {author} {\bibfnamefont {C.}~\bibnamefont {Delaere}}, \bibinfo {author} {\bibfnamefont {P.}~\bibnamefont {Demin}}, \bibinfo {author} {\bibfnamefont {A.}~\bibnamefont {Giammanco}}, \bibinfo {author} {\bibfnamefont {V.}~\bibnamefont {Lema\^\i{}tre}}, \bibinfo {author} {\bibfnamefont {A.}~\bibnamefont {Mertens}},\ and\ \bibinfo {author} {\bibfnamefont {M.}~\bibnamefont {Selvaggi}} (\bibinfo {collaboration} {DELPHES 3}),\ }\bibfield  {title} {\bibinfo {title} {{DELPHES 3, A modular framework for fast simulation of a generic collider experiment}},\ }\href {https://doi.org/10.1007/JHEP02(2014)057} {\bibfield  {journal} {\bibinfo  {journal} {JHEP}\ }\textbf {\bibinfo {volume} {02}},\ \bibinfo {pages} {057}},\ \Eprint {https://arxiv.org/abs/1307.6346} {arXiv:1307.6346 [hep-ex]} \BibitemShut {NoStop}%
\end{thebibliography}%
\end{document}